\documentclass[aps,prb,superscriptaddress,reprint,floatfix]{revtex4-1}
\usepackage{graphicx}
\usepackage{dcolumn}
\usepackage{bm}
\usepackage{amsmath}

\usepackage{color}
\usepackage{bm,graphicx,hyperref}
\hypersetup{%
  breaklinks = {true},
  citecolor = {blue},
  colorlinks = {true},
  linkcolor = {red},
}

\begin{document}

\title{Hydrogen adatoms on graphene: the role of hybridization and lattice distortion}

\author{Keian Noori}
\affiliation{Centre for Advanced 2D Materials, National University of Singapore, 6 Science Drive 2, 117546, Singapore}
	
\author{Su Ying Quek}
\affiliation{Centre for Advanced 2D Materials, National University of Singapore, 6 Science Drive 2, 117546, Singapore}
\affiliation{Department of Physics, National University of Singapore, 2 Science Drive 3, 117542, Singapore}

\author{Aleksandr Rodin}
\affiliation{Yale-NUS College, 16 College Avenue West, 138527, Singapore}
\affiliation{Centre for Advanced 2D Materials, National University of Singapore, 6 Science Drive 2, 117546, Singapore}

\date{\today}

\begin{abstract}
{Hydrogen adatoms on graphene are investigated using DFT and analytical approaches. We demonstrate that the level of lattice deformation due to the hydrogen adsorption does not substantially change the coupling between the graphene $p_z$ orbitals. The hybridization primarily takes place between the adsorbate's $s$ orbital and the graphene $p_z$ orbitals. We also show that the impurity interaction with the graphene atoms is limited to only a few nearest neighbors, allowing us to construct a compact TB model for the impurity-graphene system with an arbitrary impurity distribution. The complexity of our model scales with the number of impurities, not their separation, making it especially useful in the study of low impurity concentrations.
}
\end{abstract}

\maketitle

\section{Introduction}
\label{sec:Introduction}

Hydrogen is one of the most-studied graphene adsorbates. It has been shown, for example, that it can give rise to magnetism~\citep{Yazyev2007, Boukhvalov2008, GonzalezHerrero2016asc}, enhance  spin-orbit coupling ~\citep{CastroNeto2009}, and lead to spin relaxation~\citep{Thomsen2015} and magnetoresistance~\citep{Soriano2011}. Furthermore, the single-orbital hydrogen atom is the prototypical localized state in the problems of impurity-impurity interactions~\citep{Shytov2009lri, Agarwal2019pas, Noori2020} and charge-density oscillations.~\citep{Lawlor2013} An accurate description of the interaction between hydrogen adatoms and the underlying graphene lattice is therefore crucial to studying these important phenomena.

While density functional theory (DFT) can be used to accurately describe the effects of hydrogen adsorption, \textit{ab initio} modeling of a dilute concentration of impurities requires infeasibly large computational cells. Moreover, commonly used periodic boundary conditions cause undesirable interactions between adatoms in neighboring images (even for very large cells)~\citep{GonzalezHerrero2016asc,Noori2020}, and can induce a spurious band gap in certain cases~\citep{Garcia-Lastra2010sdb}.

A suitable tight-binding (TB) model can, in principle, overcome these limitations. When hydrogen atoms bond to graphene, they cause their host carbons to undergo a partial $sp^3$ hybridization to accommodate the adatom.~\citep{Boukhvalov2008}
The effect of this hybridization has been likened to the removal of the host's $p_z$ orbital from the lattice, creating a magnetic pseudo-vacancy.~\citep{Soriano2011, GonzalezHerrero2016asc} This approach has been successfully utilized in Ref.~\citep{Soriano2011}, where the authors used a TB mean-field Hubbard model and a supercell with hundreds of carbon atoms to calculate the system magnetization induced by this missing orbital. For two reasons, however, the vacancy model is not ideal. First, as we demonstrate in this work, the host carbon remains coupled to its neighbors, both with and without hybridization-induced lattice deformation, a fact important in the context of spin-orbit coupling. Shifting the host carbon vertically breaks the planar symmetry and leads to spin-flip scattering, whose key ingredient is the finite coupling between the $\sigma$ and $\pi$ bands of graphene.~\citep{CastroNeto2009} Additionally, even without the lattice deformation, the adatom can provide an indirect interaction between the two band families, giving rise to spin-flip processes. Treating the host site as a pseudo-vacancy therefore disallows spin scattering. Second, viewing the host atoms as vacancies precludes the treatment of the adsorbates as Anderson impurities with an on-site repulsion term $U$~\cite{Anderson1966}, thereby enforcing the use of large supercells, just as in the DFT approach. The great advantage of the Anderson model in this context is that one does not have to rely on large supercells because it is possible to integrate out the infinite graphene system and include the effects of the impurity-graphene coupling in the adatom self-energy.~\citep{Uchoa2008, Rodin2018} Such an approach leads to a much more computationally efficient implementation of the TB, especially in the case of multiple impurities, where the supercell size has to be increased even further. 

The aim of this work is to analyze the effect of impurity-induced hybridization and lattice distortion by using a combination of DFT and analytical methods to create an efficient , simplified model of adsorbates in graphene  that does not treat adatom hosts as vacancies and is compatible with the Anderson model. Graphene is described using a TB formalism, the numerical quality of which can be improved at virtually no additional computational cost (provided that all the integration is performed numerically) by extending the TB model to include more neighbors. While our analysis focuses on spin-degenerate systems with single hydrogen adsorbates, the model is general and holds for an arbitrary number of adatoms of different species and can be readily extended to include the spin degree of freedom. A major advantage of our approach is that, for multiple-impurity configurations, the computational complexity scales with the number of impurities and not their separation. This makes our model especially useful for studying the effects of impurity interactions on, for example, magnetization.

The paper is organized as follows. In Sec.~\ref{sec:Semi_hydrogenated_graphene}, we study the effects that extreme adsorption-induced lattice deformation has on the electronic properties of graphene. We observe that even a non-negligible $10^\circ$ bucking does not substantially warp the relevant bands. We also explore the semi-hydrogenated (SH) configuration using DFT and TB to understand the role that buckling plays in the adatom-graphene interaction. These observations allow us to create a simplified graphene-impurity TB model, introduced in Section~\ref{sec:Individual_Adatoms}, which is used to explore individual hydrogen adatoms on graphene. The computational methods are provided in Sec.~\ref{sec:Methods}. Finally, Sec.~\ref{sec:Conclusions} contains the concluding remarks.

\section{Semi-hydrogenated graphene}
\label{sec:Semi_hydrogenated_graphene}

\begin{figure}
    \centering
    \includegraphics[width = \columnwidth]{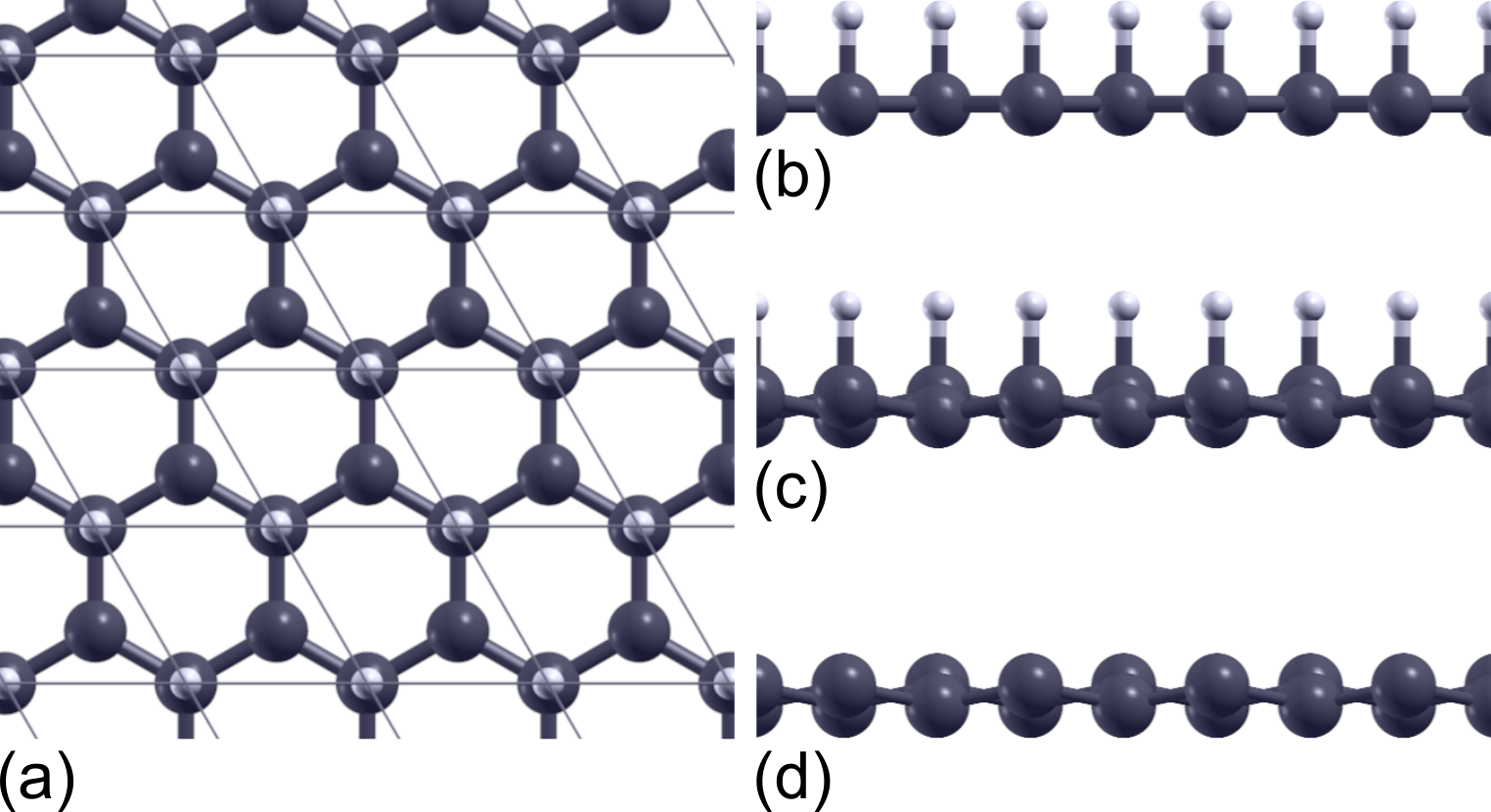}
    \caption{(a) Semi-hydrogenated planar graphene. The borders of the unit cell are drawn as gray lines. The cross sections of (b) planar and (c) structurally-relaxed SH graphene, as well as (d) buckled graphene are also shown.}
    \label{fig:structures}
\end{figure}

To illustrate the effects that hydrogen adatoms have on graphene, while isolating the role of the $sp^3$-induced buckling, we perform four band structure calculations: one for pristine graphene, two for SH graphene with and without structural relaxation, and one for buckled graphene. For the SH configurations, all the atoms of one of the graphene sublattices host a hydrogen adatom. The buckled graphene has the same lattice distortion as the relaxed SH monolayer, but without the hydrogen adatoms. The SH and buckled lattices are shown in Fig.~\ref{fig:structures}. 

\begin{figure}
    \centering
    \includegraphics[width = \columnwidth]{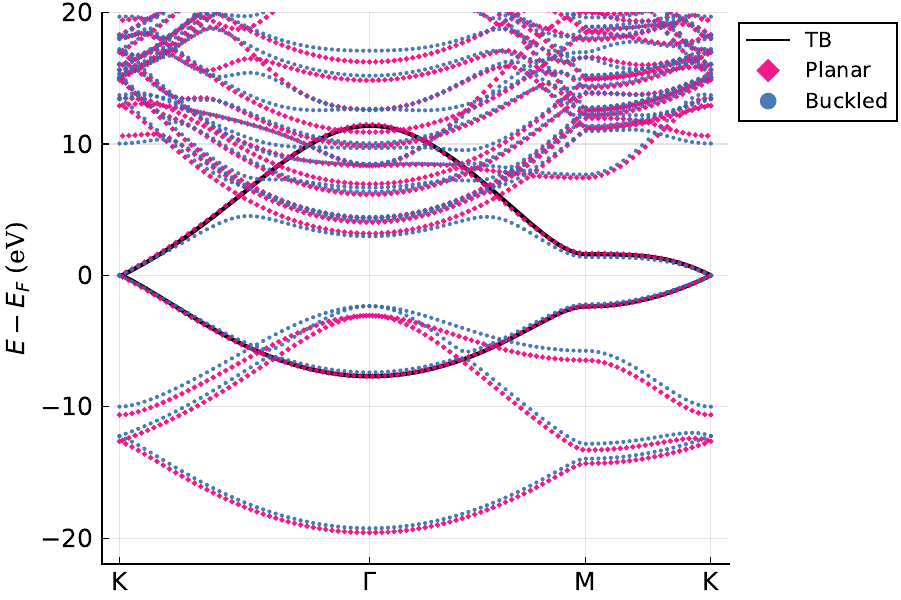}
    \caption{\emph{Ab initio} band structures for planar and buckled graphene. The solid black line is the seventeen-parameter tight-binding model from Ref.~\cite{Jung2013}.}
    \label{fig:non_H_bands}
\end{figure}

In the buckled configuration, one of the sublattices is elevated above the original graphene plane. The angle between the original plane and the bond connecting the elevated atoms to their neighbors is about $10.3^\circ$, or slightly more than half of the angle in a true $sp^3$ hybridization, where it is $19.5^\circ$. Note that when we relax the lattice, graphene is not allowed to contract laterally in response to the out-of-plane deformation because we do not expect a substantial system-wide structural modification due to individual impurities, which are the main subjects of this study. Thus, in the relaxed SH and buckled configurations, the bond length between neighboring carbon atoms becomes 1.44~{\AA} and a vertical interatomic separation is 0.26~\AA. For the sake of consistency, the hydrogen-carbon bond length for planar SH lattice is fixed to the value obtained for the relaxed SH lattice (1.18~\AA). In order to focus on the effects of buckling-induced hybridization, in the discussion that follows and in Sec.~\ref{sec:Individual_Adatoms} we have ignored the effect of electron spin. The inclusion of spin polarization does not impact our results and conclusions, and more discussion on its effects is provided in Appendix~\ref{sec:Spin}. We stress that our model can be straightforwardly extended to include the spin degree of freedom using the Anderson formalism.

Lattice buckling breaks the planar symmetry of the graphene monolayer and introduces coupling between previously isolated bands. The consequences of this interaction can be most clearly seen in Fig.~\ref{fig:non_H_bands}, which shows the comparison between the planar and buckled graphene band stuctures. We observe a strong modification of the upper branch of the buckled graphene $p_z$ band due to the mixing with the high-energy states. In addition, the new coupling creates an avoided crossing between $\pi$ and $\sigma$ valence bands at $\approx - 7$~eV, though the smallness of this level repulsion indicates that the deformation-induced interaction between $p_z$ and the other carbon valence orbitals is fairly weak.

Comparison of the band structures also reveals that the lattice deformation does not substantially modify the shape of the low-energy bands. Instead, it changes the energy separation between the $\pi$ and the $\sigma$, introducing small, rigid energy shifts in these bands with respect to their planar counterparts. A closer look reveals that the buckled bands are marginally flatter compared to the planar ones (seen best for the lower branch of the $\pi$), which can be attributed to a very modest weakening of the carbon-carbon coupling due to the bond elongation. The robustness of the band structure shape means that the adsorption-induced lattice deformation not only does not create a vacancy, but is, in fact, quite inconsequential for the coupling between the carbon valence orbitals. 

\begin{figure*}
    \centering
    \includegraphics[width = 3in]{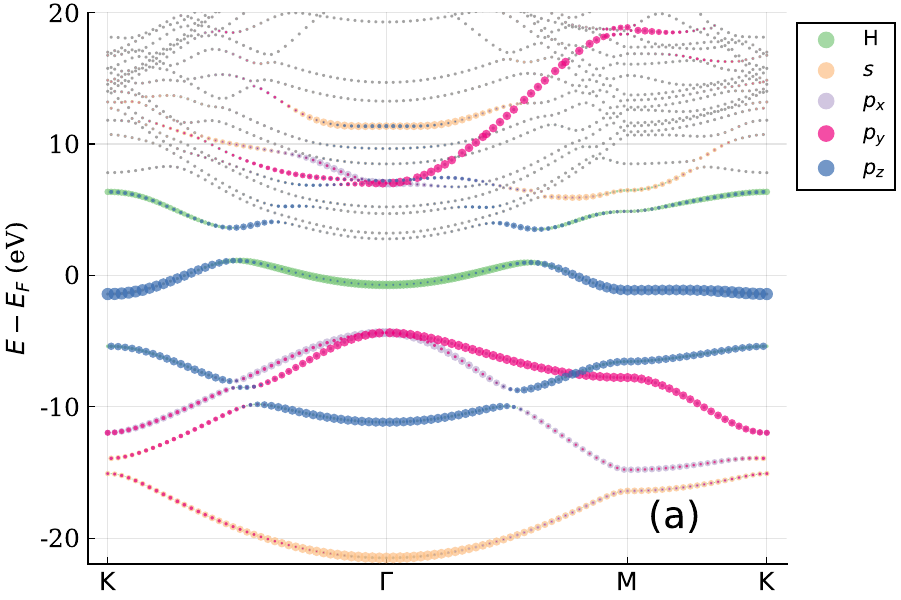}
    \includegraphics[width = 3in]{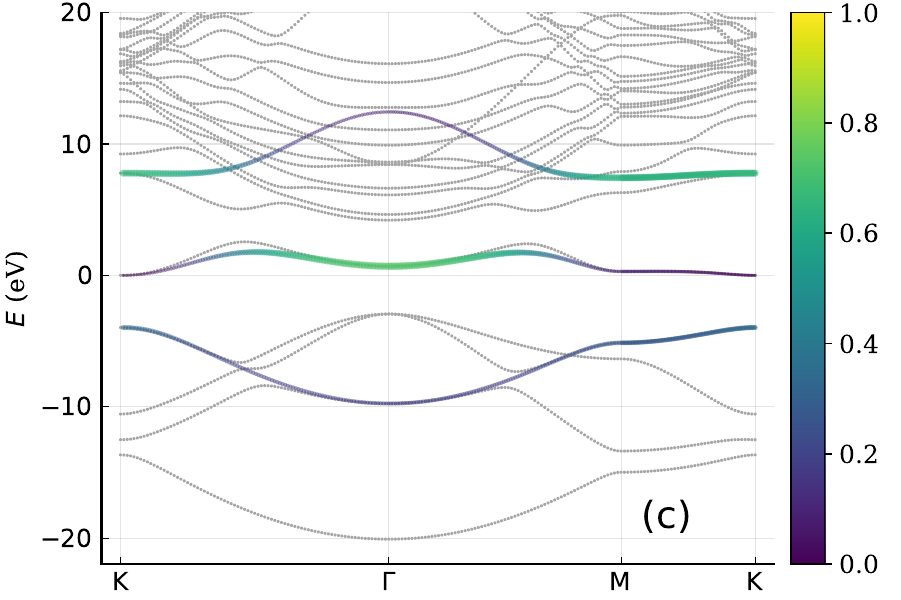}
    \includegraphics[width = 3in]{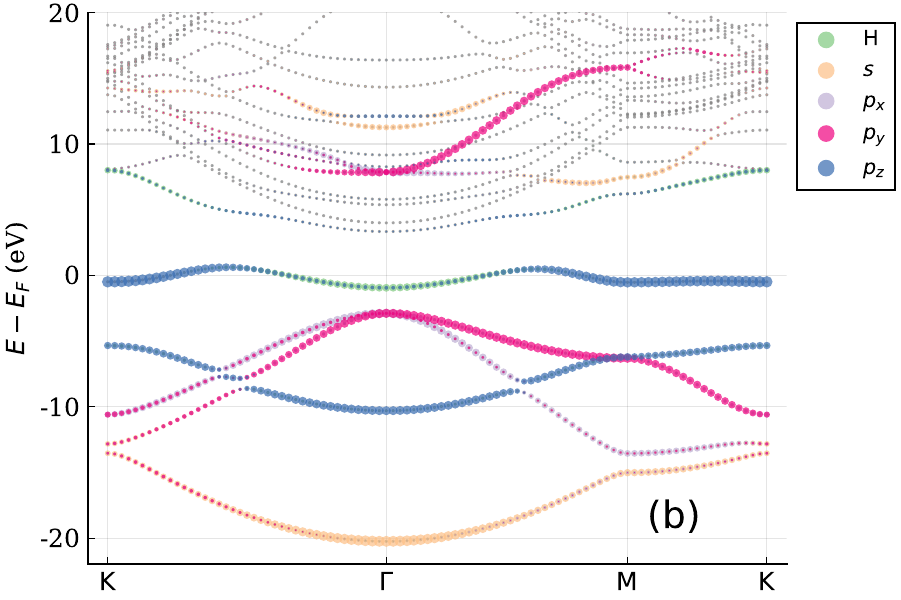}
    \includegraphics[width = 3in]{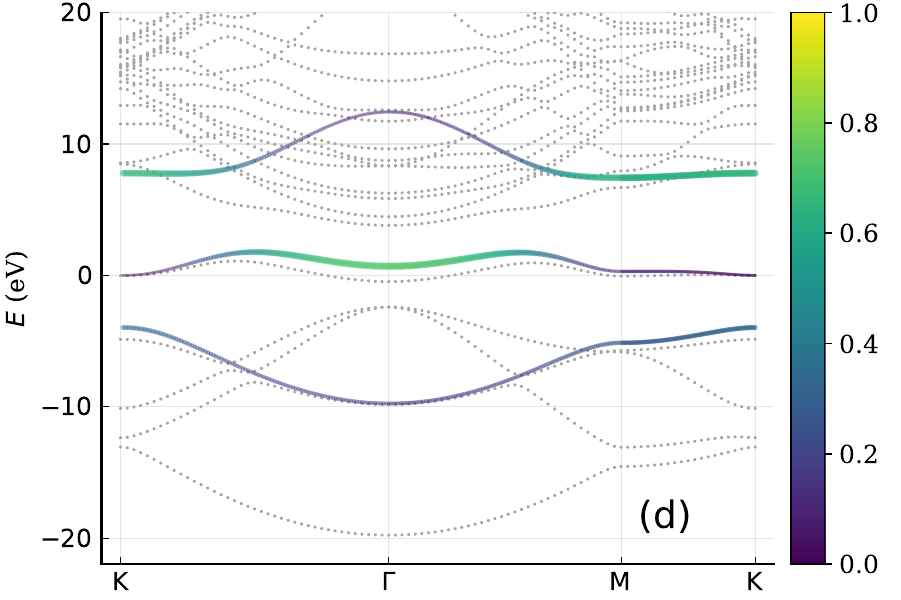}
    \caption{DFT band structures SH planar (a) and relaxed (b) graphene. The thin gray lines are all the bands, while the colored overlays illustrate the contribution of individual hydrogen and carbon valence orbitals. The size of the colored markers corresponds to the amount of a given orbital in a particular state. (c) TB fit for the $\mathrm{H}_s/\mathrm{C}_{p_z}$ bands for the flat SH lattice in panel (a) using Eq.~\eqref{eqn:H}. The color scale and the line thickness reflect the contribution of the hydrogen orbital to a given state. The upper TB band can be uncovered in (a) by following the trail of blue $p_z$ orbitals along the higher-energy bands. (d) TB approximation for the relaxed lattice in panel (b). Here, the parameters $\varepsilon$, $h$, and $V_0$ are the same as in (c), while the rest are set to zero.
    }
    \label{fig:UC_Projected}
\end{figure*}

In addition to the DFT band structures, we also plot a TB dispersion for the $\pi$ bands in Fig.~\ref{fig:non_H_bands}. This dispersion is calculated from the seventeen-parameter TB model for the $p_z$ orbitals obtained from the maximally localized Wannier functions in Ref.~\citep{Jung2013}. The TB model gives an excellent agreement with the DFT result for the pristine monolayer, as can be seen in Fig.~\ref{fig:non_H_bands}, where the solid TB curve follows the \emph{ab initio} $p_z$-orbital bands. For the buckled system, the TB bands follow the DFT results very well up to the energies where the deformation-induced mixing with higher bands becomes important. Because the TB model describes the buckled monolayer well in the energy range of interest, we will use it in our description of graphene for the relaxed SH system.

Next, we consider a SH flat lattice. Our aim here is to construct a TB model that correctly captures the coupling between the adatoms and graphene using the pristine-graphene model from Ref.~\citep{Jung2013} as the starting point. The DFT band structure is given in Fig.~\ref{fig:UC_Projected}(a), where we show the contributions of the carbon and hydrogen valence orbitals to the band composition. The hydrogen $s$ orbital sits above the graphene plane and therefore can couple to all the carbon valence orbitals. Because of this interaction, both the $\pi$ and $\sigma$ graphene bands mix with the impurity and, thus, indirectly couple to each other. The effects of this mixing can be seen as the avoided band crossings in the valence bands at $\approx -9$~eV. Curiously, the indirect hydrogen-mediated level repulsion is stronger than the direct buckling-induced one, as can be seen by comparing the avoided crossings of the buckled bands in Fig.~\ref{fig:non_H_bands} and the flat SH lattice in Fig.~\ref{fig:UC_Projected}(a). We will address this point in more detail when we consider the relaxed SH lattice.

Comparing the band structure of the flat SH system to that of pristine graphene, it is clear that the valence $\sigma$ bands remain relatively unaltered by the addition of hydrogen. The apparent downward energy shift of the SH $\sigma$ bands is a consequence of a raised Fermi level due to the electrons contributed by the hydrogen adatoms.

Unlike the $\sigma$ bands, the graphene $\pi$ bands undergo a substantial modification when the hydrogen atoms are added to system. From the band composition, it is obvious that, aside from the points of avoided crossing, the lower distorted $\pi$ bands contain primarily carbon's $p_z$ and hydrogen's $s$ orbitals. Therefore, as we construct the TB model, we ignore the  coupling of hydrogen to the non-$p_z$ carbon orbitals.

In the absence of graphene, the hydrogen atoms form a triangular lattice with an interatomic distance of 2.46~{\AA} and lattice vectors $\mathbf{d}_1$ and $\mathbf{d}_2$ identical to those of graphene. Because of the relatively large separation with respect to the size of a hydrogen atom, it is not unreasonable to assume that, from the TB perspective, the interaction strength between hydrogen atoms decreases sharply past the first nearest neighbors. Retaining only the nearest-neighbor interaction gives rise to a single energy band $E_\mathbf{q}^H = \varepsilon + h f_{2,\mathbf{q}}$, where $\varepsilon$ is the on-site energy, $h$ is the hopping term, and $f_{2,\mathbf{q}} = 2\left[\cos\left(\mathbf{d}_1\cdot\mathbf{q}\right)+\cos\left(\mathbf{d}_2\cdot\mathbf{q}\right)+\cos\left(\left(\mathbf{d}_1 - \mathbf{d}_2\right)\cdot\mathbf{q}\right)\right]$. Note that $f_{2,\mathbf{q}}$ is also the phase of the second nearest neighbor hopping in graphene. For $h < 0$, this hydrogen band has a minimum at $\Gamma$, a maximum at K, and a saddle point at M.

When graphene is introduced, hydrogen couples to its $p_z$ orbitals and any degeneracy between the $p_z$ and the hydrogen bands is lifted. In Fig.~\ref{fig:UC_Projected}(a), this level repulsion takes the form of the avoided crossing at $\approx 3$~eV between the green impurity band and the blue $\pi$ branch.

To describe the interaction between the hydrogen lattice and graphene, we write the TB Hamiltonian as
\begin{equation}
    H_\mathbf{q} = 
    \begin{pmatrix}
    \varepsilon + hf_{2,\mathbf{q}} & V_0 + V_2 f_{2,\mathbf{q}}& V_1 f_{1,\mathbf{q}} + V_3 f_{3,\mathbf{q}}
    \\
   V_0 + V_2 f_{2,\mathbf{q}} &H_\mathrm{AA}&H_\mathrm{AB}
    \\
   V_1 f_{1,\mathbf{q}}^* + V_3 f_{3,\mathbf{q}}^* &H_\mathrm{AB}^*&H_\mathrm{BB}
    \end{pmatrix}\,,
    \label{eqn:H}
\end{equation}
where $V_j$ is the coupling to the $j$th nearest neighbor of a given hydrogen atom and $f_{j,\mathbf{q}}$ is the corresponding phase factor. The bottom-right $2\times 2$ block is the seventeen-parameter graphene TB Hamiltonian.~\citep{Jung2013} Note that we modify the carbon on-site energy from Ref.~\citep{Jung2013} to ensure that the Dirac point is at $E = 0$.

There are several methods available to determine the six parameters in Eq.~\eqref{eqn:H}. One would be to perform a least-squares fitting of the bands computed from Eq.~\eqref{eqn:H} to the relevant DFT bands over the entire BZ. The main weakness of this approach in the present context comes from the fact the the upper band of Eq.~\eqref{eqn:H} lies in the range where the coupling with high-energy bands is substantial, except at certain high-symmetry points, making the fit unreliable. This can be understood from Fig.~\ref{fig:UC_Projected}(a), where the top $p_z$ is seen to mix strongly with other bands. Another approach, followed here, is to choose six energies $\lambda_k$ at high-symmetry points composed of the relevant orbitals and, by enforcing $\det\left(H_\mathbf{q} - \lambda_k\right) = 0$, obtain $\varepsilon$, $h$, $V_0$, $V_1$, $V_2$, and $V_3$. The states that we pick are the highest- and lowest-energy $\mathrm{H}_s/\mathrm{C}_{p_z}$ states at the K point and the two lowest $\mathrm{H}_s/\mathrm{C}_{p_z}$ states at the $\Gamma$ and M points. Using the lower band energies at the $\Gamma$ and M points ensures that the fit is performed on the states with the correct composition. At the K point, the middle-energy state is composed entirely of the non-host $p_z$ orbital, while the other two states are made up almost exclusively of the host $p_z$ and the adsorbate orbital. The resultant parameter values are: $\varepsilon \approx 2.44$~eV, $h \approx  -0.457$~eV, $V_0 \approx-5.55$~eV, $V_1 \approx -0.245$~eV, $V_2 \approx 0.0026$~eV, and $V_3\approx 0.0734$~eV.

The TB fit, plotted in Fig.~\ref{fig:UC_Projected}(c), shows an excellent agreement with the DFT results for the lowest band and the low-energy portion of the middle band. At higher energies, mixing with the orbitals not considered in our model becomes important. Nevertheless, it is still possible to resolve the traces of the top band in the DFT results, Fig.~\ref{fig:UC_Projected}(a), by following the states containing the $p_z$ orbitals.

Finally, we address the relaxed SH system, whose band structure is shown in Fig.~\ref{fig:UC_Projected}(b). The magnitude of the level repulsion between the valence $\pi$ and $\sigma$ bands, in this case, is more similar to the buckled graphene than to the SH flat graphene. As we noted earlier, the hydrogen-mediated splitting is greater than the deformation-induced one. Thus, the weakened repulsion in the relaxed case implies that the role of hydrogen adatoms is diminished compared to the flat lattice. This diminution is attributable to the increased distance between the adatoms and their neighboring carbon atoms.

To illustrate the effects of the reduced coupling between hydrogen and its neighboring carbons, we again turn to TB. Due to the increased distance, we set, as a first approximation, $V_{1,3} = 0$. For simplicity, we also neglect $V_2$ because of its smallness. The results of the TB calculations are shown in Fig.~\ref{fig:UC_Projected}(d). This simplified model correctly captures the flatter dispersion of the middle band and is in a fairly good agreement with the lowest band, even though the fit was calculated for a system with a substantially different structure.

\section{Individual Adatoms}
\label{sec:Individual_Adatoms}

In the section above we  established that even system-wide lattice buckling of less than $10^\circ$ does not substantially modify the low-energy graphene bands. It was also shown that the coupling between graphene and the adatoms can be described using a TB formalism with few nearest neighbors. Finally, we obtained the approximate TB energy parameters needed to describe the hydrogen-carbon coupling. Using this information, in this section we construct an analytical model for a general arrangement of adatoms. We then demonstrate the quality of this model by comparing the spectral functions obtained from it to the projected density of states (PDOS) calculated using DFT for a single hydrogen adatom. We stress that while our analysis focuses on individual hydrogen adsorbates, the formalism presented below is general and can be used for arbitrary numbers of different impurity species.

To describe an infinitely-large graphene system hosting impurity states, we use the following Hamiltonian:
\begin{align}
    \hat{H} &= \sum_{\mathbf{q}} c^\dagger_{\mathbf{q}}
    \left(H_{0,\mathbf{q}}^G - \mu\right)
    c_{\mathbf{q}}
    +
    \sum_{ k} g_{ k}^\dagger \left(\varepsilon_k - \mu\right)  g_{ k}
    \nonumber
    \\
    &+
    \sum_{ jk} 
    \left[
    c^\dagger_{\mathbf{R}_j}I_j V_{j,k} g_{ k} 
    + 
    g_{ k}^\dagger \left(V_{j,k}\right)^* I_j^T c_{\mathbf{R}_j}
    \right]
    \nonumber
    \\
    &+  
    \sum_{jl} c_{\mathbf{R}_j}^\dagger I_j \Delta_{jl} I^T_lc_{\mathbf{R}_l}
    \,.
    \label{eqn:H_QFT}
\end{align}
Here, $H_{0,\mathbf{q}}^G$ is the pristine graphene Hamiltonian matrix, $\mu$ is the chemical potential, and $c^\dagger_{\mathbf{q}} = \begin{pmatrix} a^\dagger_{\mathbf{q}}&b^\dagger_{\mathbf{q}}\end{pmatrix}$ is a vector of the creation operators for the carbon $p_z$ orbitals for the two sublattices in momentum space, while $c^\dagger_{\mathbf{R}}$ is its real-space counterpart. $g^\dagger_k$ is the creation operator for the impurity state of energy $\varepsilon_k$. The second line describes the coupling between the impurity states and graphene atoms at unit cells with coordinates $\mathbf{R}_j$. Importantly, the sum $j$ runs over \emph{all} the atoms impacted by the impurities, either by directly interacting with them or because the induced lattice deformation changes their coupling to other graphene atoms. To keep track of matrix dimensions, we denote the number of affected atoms by $M$ and the number of impurity states by $K$. The quantity $I_j^T = \begin{pmatrix} 1 & 0 \end{pmatrix}$ or $ \begin{pmatrix} 0 & 1 \end{pmatrix}$, depending on the sublattice of the atom $j$, and $V{j,k}$ is atom $j$'s interaction strength with the impurity state $k$. Finally, the last line gives the perturbation of the graphene Hamiltonian due to the lattice deformation. As with the line above, the sum includes all the modified atoms.

Before proceeding further, we highlight three important aspects of the model. First, what we refer to as the ``impurity state" is not just the adatom orbital. Rather, as a consequence of hybridization, it also includes contributions from graphene orbitals not included in the model. As a consequence, the energy $\varepsilon_k$ can depend on the carbon-adatom bond length, among other factors, as it influences the magnitude of the orbital interaction. Second, the impurity states are not orthogonal to the graphene Wannier functions due to a finite overlap integral. Equation~\eqref{eqn:H_QFT} assumes that the overlap is small and neglects it by treating all the states in the system as orthogonal. While it is possible to extend the treatment to non-orthogonal states, this is outside the scope of our paper. Finally, even though the Hamiltonian in Eq.~\eqref{eqn:H_QFT} includes only the carbon $p_z$ orbitals, the subsequent derivation does not depend on this fact. Put differently, to include more orbitals in the model, one simply needs to modify $H_{0,\mathbf{q}}^G$ and adjust the dimensions of $I_j$ in the final result. In this case, the $j$ and $l$ summations run over \emph{orbitals}, not \emph{atoms}.

Using  $c^\dagger_\mathbf{R} = N^{-1/2}\sum_\mathbf{q}c^\dagger_\mathbf{q} e^{-i\mathbf{R}\cdot\mathbf{q}}$, where $N$ is the number of unit cells in the system, we write
\begin{equation}
    \sum_{j}
    c^\dagger_{\mathbf{R}_j} I_j V_{j,k}
    = 
    \frac{1}{\sqrt{N}}
    \sum_\mathbf{q}c^\dagger_{\mathbf{q}}
    \underbrace{\left(\sum_{j}e^{-i\mathbf{R}_j\cdot\mathbf{q}} I_j V_{j,k}\right)}_{\Theta_\mathbf{q}^\dagger \mathbf{I} V_{,k}}\,,
    \label{eqn:Coupling}
\end{equation}
where $\Theta_\mathbf{q}$ is a column vector of $\mathbf{1}_{2\times 2} e^{i\mathbf{R}_j\cdot\mathbf{q}}$ for all $\mathbf{R}_j$, $\mathbf{I}$ is a diagonal matrix of $I_j$, and $V_{,k}$ is a column vector of $V_{j,k}$. Similarly,
\begin{align}
	\sum_{jk} c_{\mathbf{R}_j}^\dagger I_j \Delta_{jk} I^T_kc_{\mathbf{R}_k}
	=\frac{1}{N}\sum_{\mathbf{qq}'} c_{\mathbf{q}}^\dagger \Theta_{\mathbf{q}}^\dagger\mathbf{I}\Delta \mathbf{I}^T\Theta_{\mathbf{q}'}c_{\mathbf{q}'}\,,
	\label{eqn:Delta}
\end{align}
where $\Delta$ is an $M\times M$ matrix.

Plugging Eqs.~\eqref{eqn:Coupling} and \eqref{eqn:Delta} into Eq.~\eqref{eqn:H_QFT}, we transcribe the Hamiltonian into the imaginary time action
\begin{align}
    S &= \sum_{\omega_n\mathbf{qq}'} \bar\psi_{\omega_n\mathbf{q}}
    \overbrace{
    \left[\left(-i\omega_n-\mu\right) \delta_{\mathbf{qq}'}
    + 
    H^G_{\mathbf{qq}'} \right]}^{-G^{-1}_{i\omega_n + \mu, \mathbf{qq}'}}
    \psi_{\omega_n\mathbf{q}'}
    \nonumber
    \\
    &
    +
    \sum_{\omega_n k}\bar\gamma_{\omega_n,k} 
    \overbrace{\left(-i\omega_n - \mu + \varepsilon_k \right)}^{-\Gamma_{0,i\omega_n + \mu, k}^{-1}} \gamma_{\omega_n,k}
    \nonumber
    \\
    &
    +
    \frac{1}{\sqrt{N}}\sum_{\omega_n k \mathbf{q}} 
    \left(
    \bar\psi_{\omega_n,\mathbf{q}}
    \Theta_\mathbf{q}^\dagger
    \mathbf{I}V_{,k}
    \gamma_{\omega_n,k}
    \right.
    \nonumber
    \\
    &\quad\quad\quad\quad\quad
    +
    \left.
    \bar\gamma_{\omega_n,k} 
    V_{,k}^T\mathbf{I}^T\Theta_\mathbf{q}
    \psi_{\omega_n,\mathbf{q}}
    \right)\,.
    \label{eqn:S}
\end{align}
Note that we have combined the $\mathbf{q}$-diagonal and non-diagonal portions of the graphene Hamiltonian into $ H^G_{\mathbf{qq}'}$. The quantity $\omega_n$ is the fermionic Matsubara frequency, and $\gamma$ and $\psi$ are Grassmann fields. Integrating $e^{-S}$ over all the fields yields the partition function
\begin{widetext}
\begin{align}
    \mathcal{Z} &= 
    \prod_{\omega_n }
    \left|-\beta G^{-1}_{i\omega_n+\mu}\right|
    \left|-\beta\left(\Gamma^{-1}_{0,i\omega_n+\mu} - \frac{V^T \mathbf{I}^T\Theta  G_{i\omega_n+\mu} \Theta^\dagger \mathbf{I} V}{N}\right)\right|
    \nonumber
    \\
    &= 
    \prod_{\omega_n }
    \left|-\beta \Gamma^{-1}_{0,i\omega_n+\mu}\right|
    \left|-\beta\left( G_{i\omega_n+\mu} ^{-1} -  \frac{\Theta^\dagger\mathbf{I}V\Gamma_{0,i\omega_n+\mu}V^T \mathbf{I}^T\Theta  }{N}\right)\right|\,,
    \label{eqn:Z}
\end{align}
\end{widetext}
where $\Theta$ as a row vector of $\Theta_\mathbf{q}$ and $V$ is an $M\times K$-dimensional matrix. Defining a pristine graphene Green's function $G_{z}^0 = \left(z - H^G_0\right)^{-1}$, we get
\begin{align}
	G_{z} &= \left[\left(G_{z}^0\right)^{-1} - \frac{1}{N} \Theta^\dagger\mathbf{I}\Delta \mathbf{I}^T\Theta\right]^{-1}
	\label{eqn:G}
	\\
	&= 
	G_{z}^0
	+
	\frac{1}{N} G_{z}^0\Theta^\dagger\mathbf{I}\Delta
	\left( 1
	-
	\mathbf{I}^T\boldsymbol{\Xi}_{z}\mathbf{I}\Delta 
	\right)^{-1}\mathbf{I}^T\Theta G_{z}^0\,,	
	\nonumber
\end{align}
where $\boldsymbol{\Xi}_{z} = \Theta G_{z}^0\Theta^\dagger / N$ with entries $\boldsymbol{\Xi}_{z}^{jk} = \Xi_{z}^{\mathbf{R}_j - \mathbf{R}_k}$ and
\begin{equation}
    \Xi_z^\mathbf{R} = \frac{1}{N}\sum_\mathbf{q}G_{z\mathbf{q}}^0
    e^{i \mathbf{R} \cdot\mathbf{q}}\,.
    \label{eqn:Xi}
\end{equation}
$G_z$ is the graphene Green's function including the lattice deformation, but not the effects of the impurity states.

In the parentheses of the first line of Eq.~\eqref{eqn:Z}, we identify the inverse of the full impurity Green's function, denoted by $\Gamma^{-1}_{z}$:
\begin{align}
	\Gamma_{z} & = \left(
    \Gamma^{-1}_{0,z}
    - 
   V^T\Lambda_{z}V
	\right)^{-1}
	\nonumber
	\\
	&= 
	\Gamma_{0,z}
	+
	\Gamma_{0,z}
    V^T\Lambda_{z}
	\left(
	1
     - V
    \Gamma_{0,z}
    V^T\Lambda_{z}
	\right)^{-1}V\Gamma_{0,z}
	\,,
    \label{eqn:Gamma}
    \\
    \Lambda_{z} &=
    \mathbf{I}^T\boldsymbol{\Xi}_{z}\mathbf{I}
	\left[1+
    \Delta
	\left( 1
	-
	\mathbf{I}^T\boldsymbol{\Xi}_{z}\mathbf{I}\Delta 
	\right)^{-1} \mathbf{I}^T\boldsymbol{\Xi}_{z}\mathbf{I}
	\right]\,.
	\label{eqn:Lambda}
\end{align}

Also, from the parentheses of the second line in Eq.~\eqref{eqn:Z}, we obtain the inverse of the full graphene Green's function, given by
\begin{align}
	 \mathcal{G}_{z} & =\left[\left(G_{z}^0\right)^{-1} - \frac{1}{N} \Theta^\dagger\mathbf{I}\left(\Delta + V\Gamma_{0,z}V^T
	 \right)\mathbf{I}^T\Theta
	 \right]^{-1}
	 \nonumber
	 \\
	 &=
	 G_{z}^0+
	  \frac{1}{N} G_{z}^0\Theta^\dagger\mathbf{I} 
	 D_{z}
	 \mathbf{I}^T\Theta
	 G_{z}^0\,,
	 \label{eqn:Full_G}
	 \\
	 D_{z} &=\left[
	\left(\Delta + V\Gamma_{0,z}V^T
	 \right)^{-1}
	 -
	\mathbf{I}^T
	\boldsymbol{\Xi}_{z} \mathbf{I}
	 \right]^{-1}\,.
	 \label{eqn:D}
\end{align}

Using Eq.~\eqref{eqn:Full_G}, it is possible to calculate the real-space graphene Green's function $\mathcal{G}_{i\omega_n + \mu,\mathbf{R}}^s = N^{-1}\sum_{\mathbf{qq}'} \langle \bar{\psi}^s_{\omega_n\mathbf{q}}\psi^s_{\omega_n\mathbf{q}'}\rangle e^{i\left(\mathbf{q}'-\mathbf{q}\right)\cdot\mathbf{R}}$, where $s$ denotes the sublattice and the correlation functions are the diagonal elements of the $\left[\mathcal{G}_{i\omega_n+\mu} \right]_{\mathbf{q}'\mathbf{q}}$ blocks:
\begin{align}
	\mathcal{G}_{z,\mathbf{R}} 
	 &=\Xi^\mathbf{0}_{z}
	 +
	\sum_{jk}
	\Xi_{z}^{\mathbf{R} - \mathbf{R}_j}
	  \left(\mathbf{I} 
	 D_{z}
	 \mathbf{I}^T
	 \right)_{jk}
	 \Xi_{z}^{\mathbf{R}_k - \mathbf{R}}
	 \label{eqn:G_R}
	 \\
	 &=\Xi^\mathbf{0}_{z}
	 +
	 \begin{pmatrix}
	 	\Xi_{z}^{\mathbf{R} - \mathbf{R}_1} & \cdots
	 \end{pmatrix}
	 \mathbf{I} 
	 D_{z}
	 \mathbf{I}^T
	 \begin{pmatrix}
	 	\Xi_{z}^{\mathbf{R}_1 - \mathbf{R}} \\ \vdots
	 \end{pmatrix}\,.
	 	 \nonumber
\end{align}

By taking the $k$th diagonal entry of $-2\mathrm{Im}\left[\Gamma_{\omega + i0^+}\right]$ and $-2\mathrm{Im}\left[\mathcal{G}_{\omega+i0^+,\mathbf{R}}^s\right]$, we obtain the spectral functions for the $k$th impurity and the corresponding carbon atom, respectively. These spectral functions can be compared directly to the DFT-computed PDOS.

To model individual hydrogen atoms using DFT, we increase the size of the graphene supercell hosting a single impurity to $10\times10$ unit cells. As before, we perform calculations using both planar and relaxed graphene while setting the carbon-hydrogen bond length to the value obtained for the relaxed configuration (1.13~\AA). For the relaxed lattice, the bond length between the host carbon and its nearest neighbors is 1.50~{\AA} with a vertical distance of 0.35~{\AA}. The first neighbors, on the other hand, sit about 0.15~{\AA} above the original plane (close to 0.08~{\AA} above the second neighbors). As one goes farther from the impurity, the height difference between neighbors continues to decrease. We saw in the case of SH graphene that the vertical displacement of 0.25~{\AA} does not drastically impact the band structure. In the present case, since, except for the host atom, all neighbors have a relative vertical displacement that is substantially less than the SH value, it is reasonable to treat the rest of the lattice as flat.

\begin{figure}
    \centering
    \includegraphics[width = 3in]{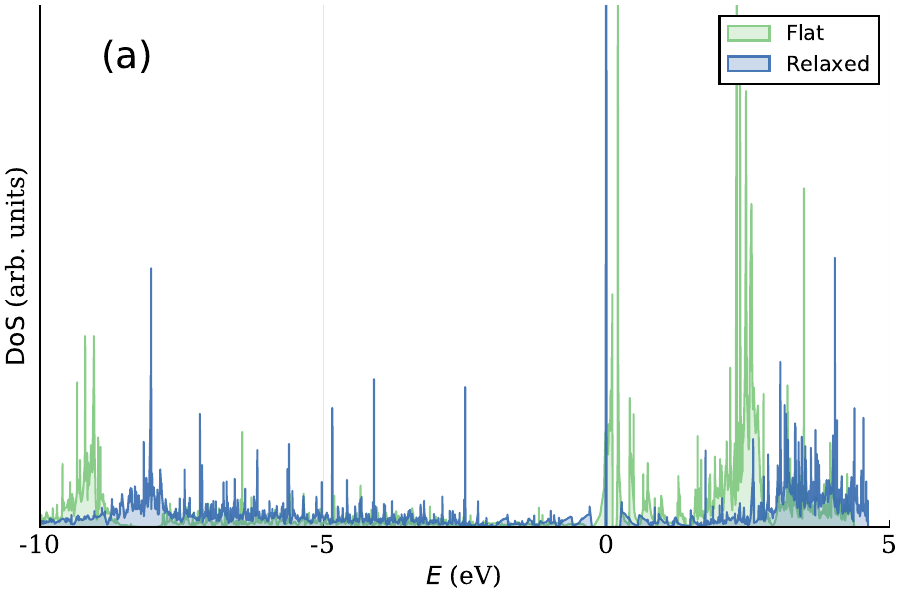}
    \includegraphics[width = 3in]{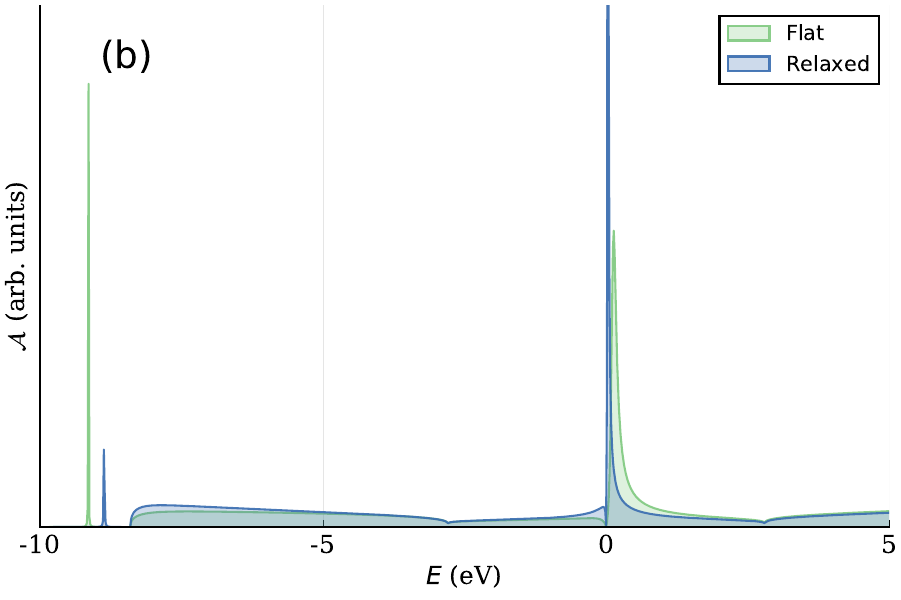}
    \caption{(a) PDOS of the hydrogen $s$ orbital for a $10\times 10$ unit cell. (b) Spectral function of the impurity state.}
    \label{fig:Impurity_PDOS}
\end{figure}

The PDOS for the hydrogen $s$ orbital is given in Fig.~\ref{fig:Impurity_PDOS}(a). While qualitatively similar, the flat and the relaxed results exhibit some differences. First, the peak around $E = 0$ is substantially broader for the flat configuration. Second, the broad low-energy peak for the flat PDOS is below the relaxed PDOS one. One can also see a small gap around $E = 0$ with a midgap peak for the relaxed structure. This is the spurious gap referred to in the introduction, and is the consequence of cross-supercell impurity interactions caused by periodic boundary conditions.

For the spectral function, we need to calculate $\Xi_z^\mathbf{R}$ in Eq.~\eqref{eqn:Xi}, which requires $H_{0,\mathbf{q}}^G$. In the discussion above, we used a seventeen-parameter tight-binding Hamiltonian. While it is also possible to do so here, the necessary two-dimensional integrals significantly increase the numerical cost without substantially changing the qualitative picture. Therefore, we use the nearest-neighbor hopping Hamiltonian as it allows us to take one of the momentum integrals analytically. The loss of accuracy is not a major concern here since we expect inaccuracies in the results for high-energy states even for the more elaborate model because of the mixing with other bands. Furthermore, the minimal nearest-neighbor model used here captures the qualitative behavior of the low-energy states reasonably well. We refer the reader to Appendix~\ref{sec:Propagator} for the calculation of $\Xi_z^\mathbf{R}$.

\begin{figure*}
    \centering
    \includegraphics[width = 3in]{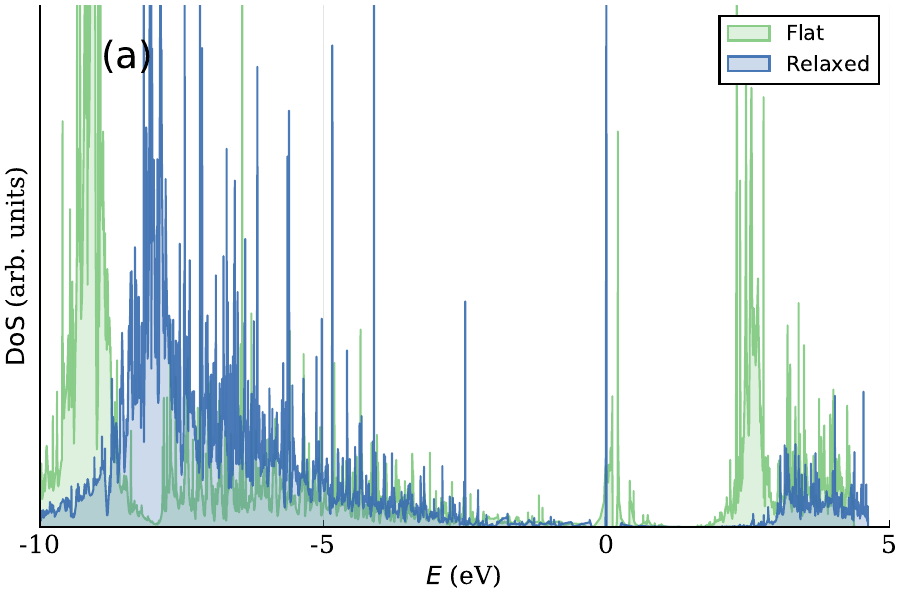}
    \includegraphics[width = 3in]{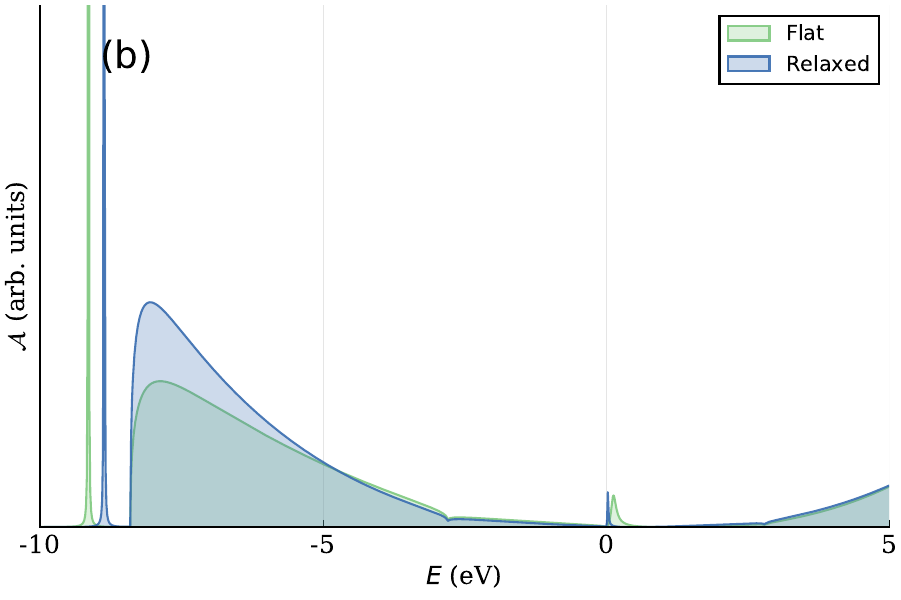}
    \\
    \includegraphics[width = 3in]{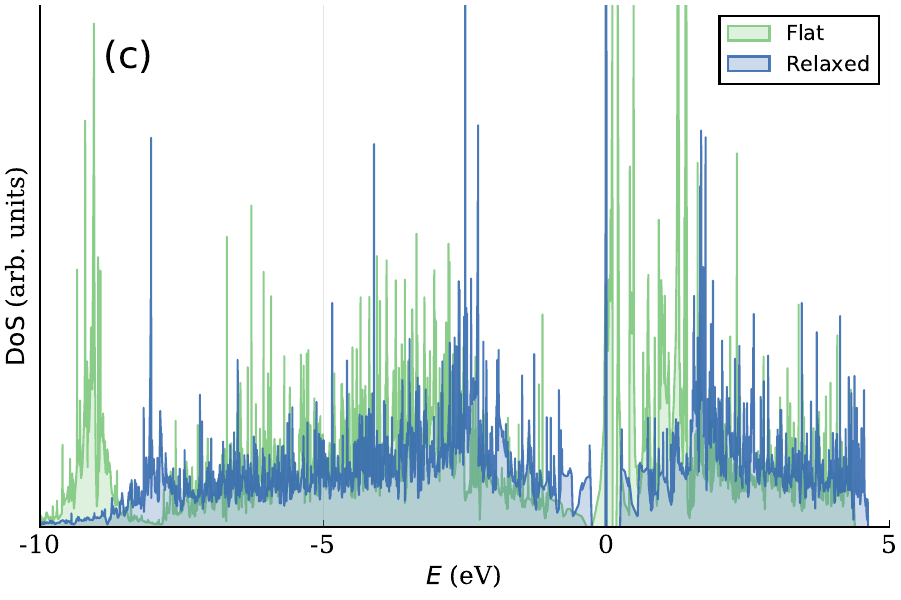}
    \includegraphics[width = 3in]{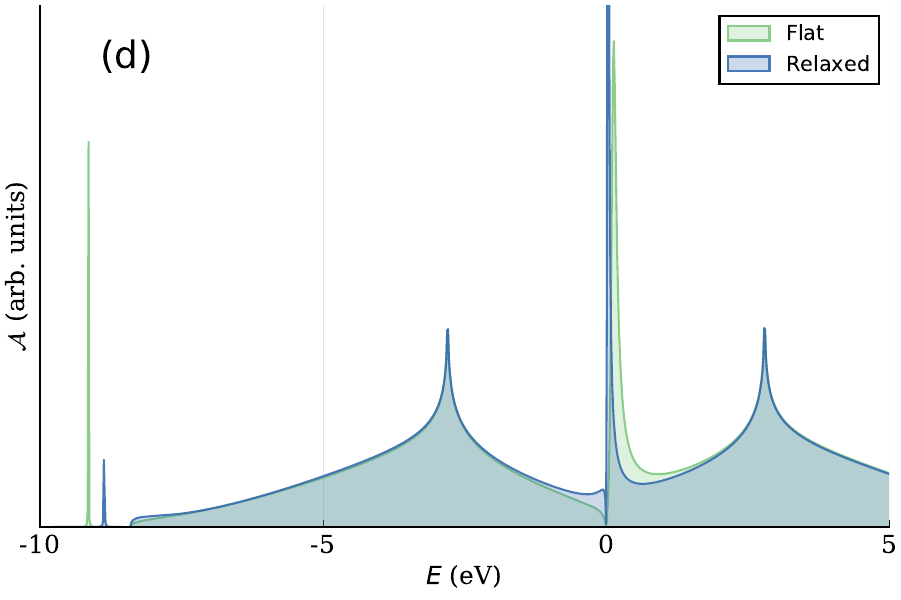}
    \\
    \includegraphics[width = 3in]{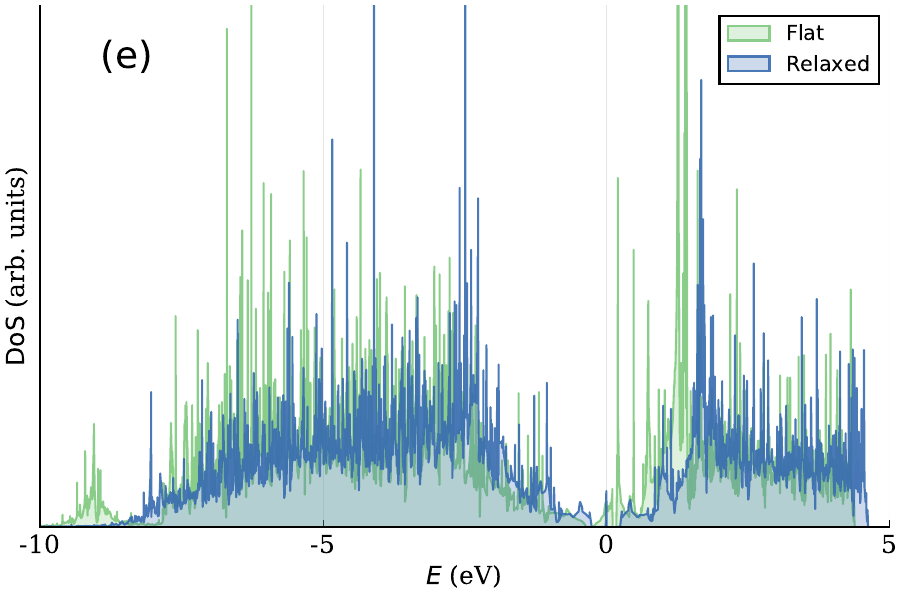}
    \includegraphics[width = 3in]{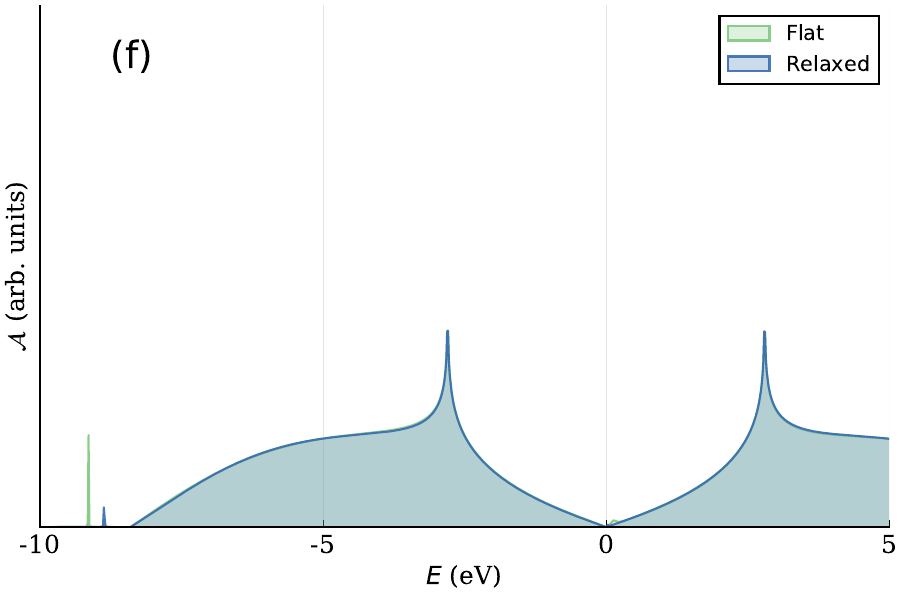}
    \\
    \includegraphics[width = 3in]{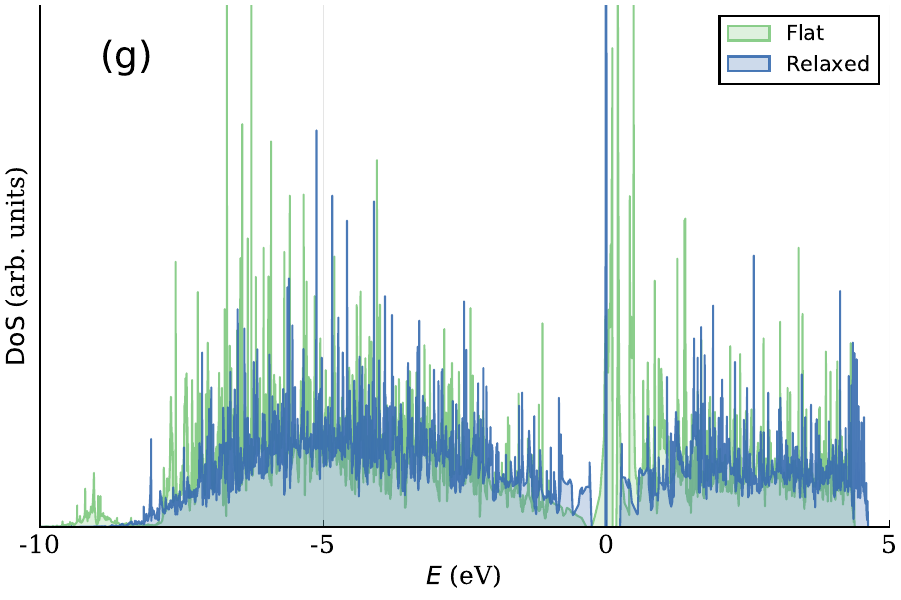}
    \includegraphics[width = 3in]{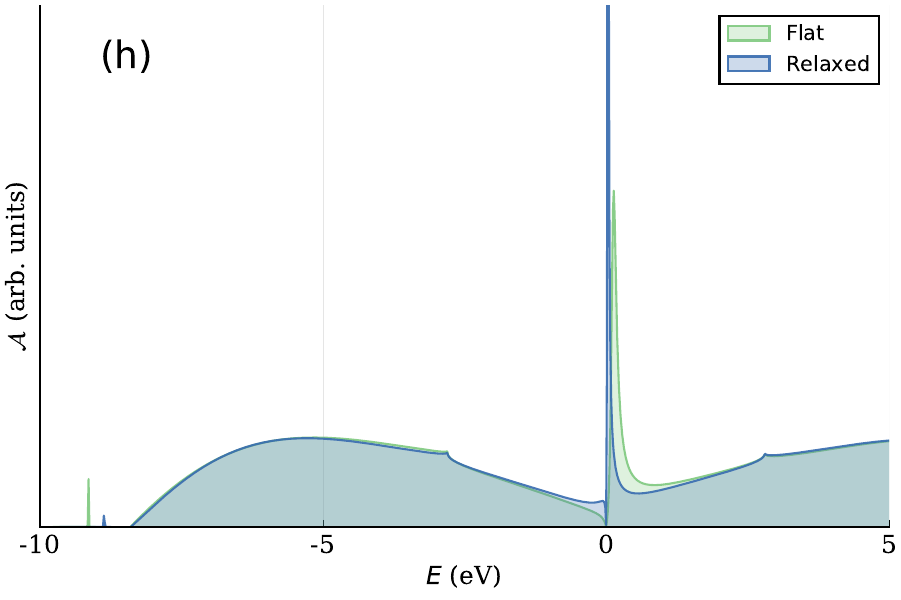}
    \caption{PDOS (left column) and the spectral function (right column) for the $p_z$ orbital of the host carbon (top row), and of the first, second, and third nearest neighbors (second, third, and fourth rows).}
    \label{fig:PDOS_10x10}
\end{figure*}

To make use of Eqs.~\eqref{eqn:Gamma} and ~\eqref{eqn:G_R}, we assume that the hydrogen adatom can interact with four carbon atoms (the host and its nearest neighbors) and that the hopping in graphene is modified only between the host and its neighbors. The motivation behind this assumption is to produce the minimal model that exhibits the principal features. The treatment is easily extendable to include more neighbors if greater numerical accuracy is desired. We use the information obtained in the previous section to make reasonable choices for the single-impurity TB parameters. For the hydrogen-carbon coupling, we take into account the reduced carbon-hydrogen bond length, as compared to the SH system, and use $V_0 = -7$~eV. To highlight the difference between the flat and the relaxed lattices, we set $V_1 = -0.2$~eV for the flat configuration and $V_1 = 0$ for the relaxed one, similar to the quantities we used for the SH graphene systems. We also reduce the hopping energy between the host and its neighbors by 5\% for the relaxed system. Finally, we set $\varepsilon = 0.5$~eV. This value is smaller than what we obtained in the SH case and is guided by the position of the PDOS peak. We stress that the qualitative behavior of the spectral function is quite insensitive to the exact parameter choice.

The spectral function for the impurity, shown in Fig.~\ref{fig:Impurity_PDOS}(b), agrees well with the impurity PDOS in the energy range where the mixing with higher-energy bands can be neglected. Below the $\pi$-band range, the spectral function contains a pole corresponding to a localized state. If $\sigma$ bands are included, this localized state mixes with them and broadens the peak, which is precisely what we observe in the PDOS.

We also calculate the PDOS and spectral function for the host carbon, as well as its first, second, and third nearest neighbors. The results are shown in Fig.~\ref{fig:PDOS_10x10}, where it is clear that the salient features are preserved across all the panels. To improve the quantitative agreement between the two types of calculations, one can use a more complete TB model for the spectral function. This would bring the van Hove singularities closer to the Dirac point and produce a particle-hole asymmetry, as observed in the DFT PDOS of panels (c) and (e). Another source of differences between PDOS and the spectral function is the presence of additional bands in the DFT calculations, giving rise to extra features at low and high energies. A more complete Hamiltonian with additional orbitals and more hydrogen-carbon coupling terms is expected to produce a better agreement, especially if the non-orthogonality of the impurity state with respect to the graphene orbitals is taken into account.

\section{Methods}
\label{sec:Methods}
DFT calculations were performed with \textsc{Quantum} ESPRESSO~\citep{Giannozzi2009,Giannozzi2017} using a projector augmented wave (PAW)~\citep{Blochl1994paw,DalCorso2014} basis and the PBE~\citep{Perdew1997gga} exchange correlation functional. The kinetic energy cutoff of wavefunctions was set to 60~Ry. Structural relaxations were performed until the forces on all atoms were below 10~meV/\AA\ and the energy difference between subsequent relaxation steps was below 0.0001~Ry. For band structure calculations, the charge densities of the systems were computed by sampling the Brillouin zones via unit cell--equivalent uniform meshes of $36 \times 36$ k-points. DOS calculations were performed using the tetrahedron method~\citep{Blochl1994a}, with unit cell--equivalent Brillouin zone samplings of $60 \times 60$ k-points (charge density) and $120 \times 120$ k-points (eigenvalues).

Numerical calculations of the model were performed using the JULIA programming language.~\citep{Bezanson2017}. The code and the DFT data can be found at https://github.com/rodin-physics/graphene-hydrogen.

\section{Conclusions}
\label{sec:Conclusions}
In summary, we have demonstrated a tight-binding model for hydrogen adsorbates on graphene. This model scales with the number of impurities and gives an excellent qualitative agreement with the DFT band structure and PDOS, even for graphene with impurity-induced lattice buckling. Our model can be trivially extended to improve numerical accuracy by including more neighbours, and can be generalized to account for multiple impurities as well as other types of impurity. Unlike existing TB models and \textit{ab initio} methods, our approach does not require large supercells, allowing us to treat an arbitary arrangement of mutliple impurities in a computationally efficient manner. We imagine that this model can be applied to treating, for example, multiple impurity-induced magnetism in graphene. 

\acknowledgements
The authors acknowledge the National Research Foundation, Prime Minister's Office, Singapore, under its Medium Sized Centre Programme, and the Ministry of Education, Singapore via grant MOE2017-T2-2-139. A. R. acknowledges support by Yale-NUS College (through grant number R-607-265-380-121). K. N. and S.Y. Q. acknowledge support from Grant NRF-NRFF2013-07 from the National Research Foundation, Singapore.

\appendix

\begin{figure}
    \centering
    \includegraphics[width = 3in]{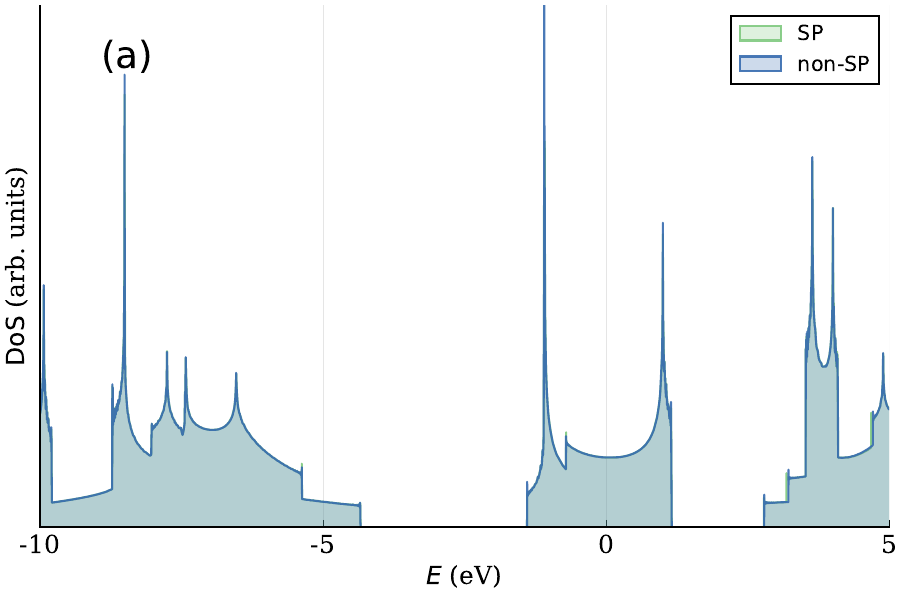}
    \\
    \includegraphics[width = 3in]{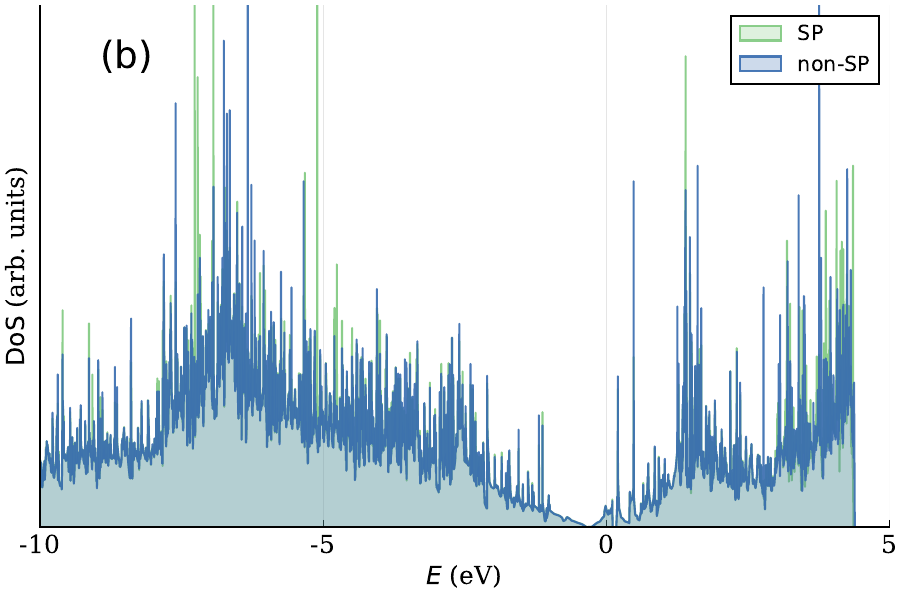}
    \caption{Density of states of the planar SH cell (a) and $10\times 10$ supercell (b) for spin-polarized and non-spin-polarized calculations.}
    \label{fig:Unrelaxed_DOS}
\end{figure}

\section{Effect of spin polarization}
\label{sec:Spin}
\begin{figure}
    \centering
    \includegraphics[width = 3in]{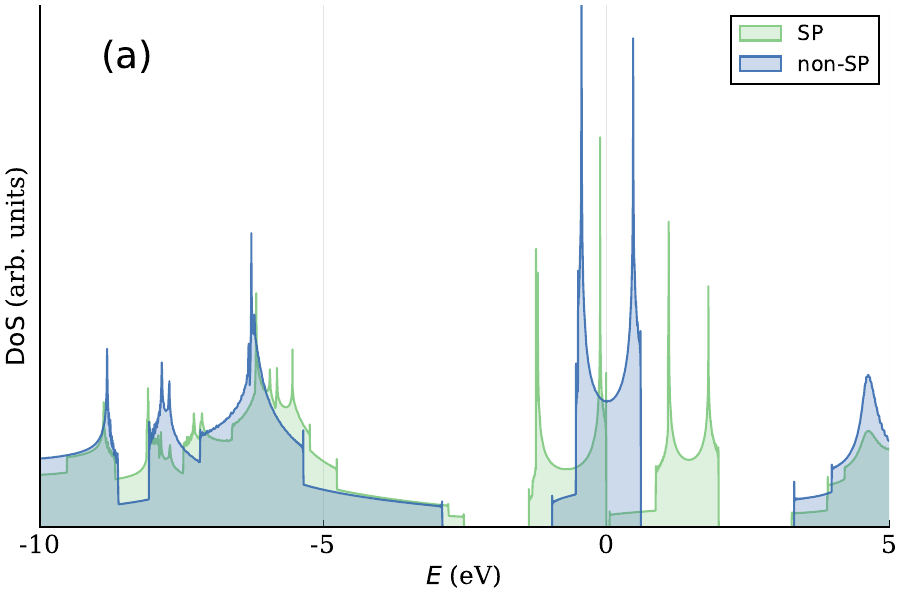}
    \\
    \includegraphics[width = 3in]{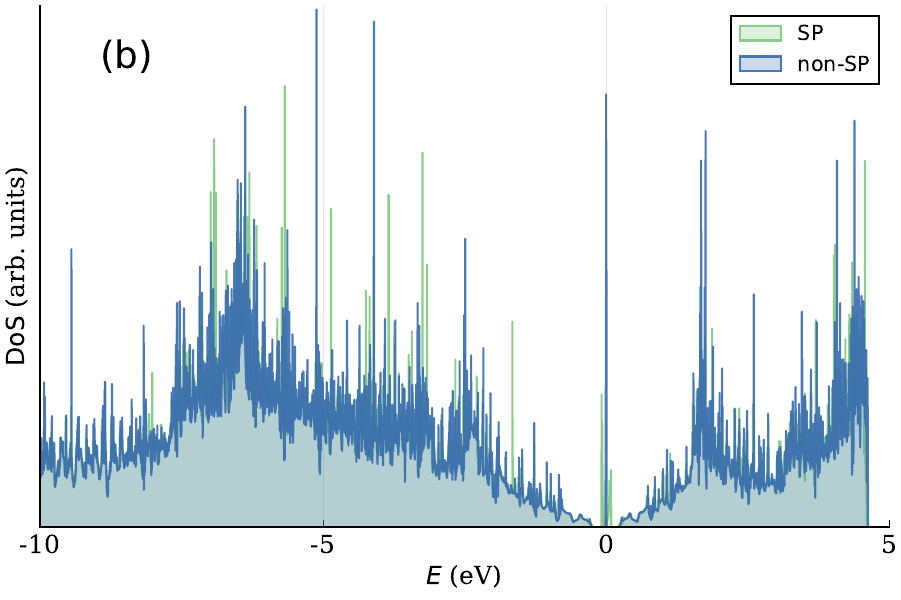}
    \caption{Density of states of the relaxed SH cell (a) and $10\times 10$ supercell (b) for spin-polarized and non-spin-polarized calculations.}
    \label{fig:Relaxed_DOS}
\end{figure}

All \textit{ab initio} electronic structure calculations were performed without spin polarization (i.e. the explicit inclusion of both spin up and down electrons in the description of the electronic density). The addition of a single hydrogen atom on graphene adds an unpaired electron that, in principle, necessitates the use spin polarized DFT calculation. In practice, however, the effect of spin polarization on the electronic structure of the planar systems examined in this work is minimal, as can be seen in Fig.~\ref{fig:Unrelaxed_DOS}. Figure~\ref{fig:Unrelaxed_DOS} shows the DOS of the planar SH cell and $10 \times 10$ supercell, both with and without spin polarization, from which we can see that the impact of spin polarization is minimal on the total DOS.  In the case of relaxed graphene (Fig.~\ref{fig:Relaxed_DOS}), the spin-polarized DOS of SH graphene and the $10 \times 10$ supercell demonstrate a clear peak splitting near the Fermi level which is absent in the non-spin polarized cases and which leads to a finite magnetic moment. The exact mechanism of this buckling-induced magnetism is beyond the scope of this discussion; here we stress that, aside from the region of the Fermi level, the overall dispersion is largely unaffected by the inclusion of spin and can therefore be safely ignored in the discussion above.

\section{Graphene Propagator}
\label{sec:Propagator}

To compute $\Xi_z^\mathbf{R}$, we first introduce
\begin{equation}
	\Omega^{u,v}_z =
	\frac{1}{N}\sum_{\mathbf{q}\in\mathrm{BZ}}
	\frac{
		e^{i\mathbf{q}\cdot \left(u\mathbf{d}_1 + v\mathbf{d}_2\right)}
	}
	{z^2 - t^2\left| f_{1,\mathbf{q}}\right|^2}
	\label{eqn:Omega_R}
\end{equation}
with $ u\mathbf{d}_1 + v\mathbf{d}_2 = \frac{d}{2}\left(u - v, \sqrt{3}\left(u+v\right)\right)$ and $t = 2.8$~eV as the nearest-neighbor hopping energy. Using $\mathbf{q}\cdot \left(u\mathbf{d}_1 + v\mathbf{d}_2\right)  = \frac{d}{2}\left[\left(u - v\right)q_x + \sqrt{3}\left(u+v\right)q_y\right]$ and turning the momentum sum into an integral yields
\begin{align}
	\Omega^{u,v}\left(z\right) 
	& = \frac{1}{\left(2\pi\right)^2}\oint dx \oint dy
	\nonumber
	\\
	&\times
	\frac
	{e^{i \left[\left(u - v\right)x + \left(u+v\right)y\right]}}
	{z^2 - t^2\left(1 + 4\cos^2 x + 4 \cos x\cos y \right)}\,.
	\label{eqn:Omega_R_2}
\end{align}
From
\begin{equation}
	\oint d\theta \frac{e^{il\theta}}{w-\cos\theta} = 2\pi \frac{\left(w - \sqrt{w - 1}\sqrt{w + 1}\right)^{|l|}}{\sqrt{w - 1}\sqrt{w + 1}}\,,
	\label{eqn:Ang_Int}
\end{equation}
we get
\begin{align}
	\Omega^{u,v}_z &= \frac{1}{2\pi}\frac{1}{4t^2}
	\oint dx \frac{e^{i\left(u - v\right)x}}{\cos x}\frac{\left(W - \sqrt{W - 1}\sqrt{W + 1}\right)^{|u+v|}}{\sqrt{W - 1}\sqrt{W + 1}}\,,
	\label{eqn:Omega_R_3}
	\\
	W &= \frac{\frac{z^2}{t^2}-1}{4\cos x}-\cos x\,.
	\label{eqn:W}
\end{align}
Finally, $\Xi^{\mathbf{R}}_z$ for $\mathbf{R} = u\mathbf{d}_1 + v\mathbf{d}_2$ can be written as
\begin{align}
	\Xi^{\mathbf{R}}_z 
	&=
	\begin{pmatrix}
		z\Omega^{u,v}_z
		&
		- t\left[\Omega^{u,v}_z + \Omega^{u,v}_{+,z} \right]
		\\
		- t\left[\Omega^{u,v}_z + \Omega^{u,v}_{-,z}\right]
		&
		z\Omega^{u,v}_z
	\end{pmatrix}\,,
	\\
	\Omega^{u,v}_{\pm,z}
	&= 
	 \frac{1}{2\pi}\frac{1}{4t^2}
	\oint dx 
	\nonumber
	\\
	&\times2e^{i\left(u - v\right)x}\frac{\left(W - \sqrt{W - 1}\sqrt{W + 1}\right)^{|u+v\pm 1|}}{\sqrt{W - 1}\sqrt{W + 1}}
\,.
\end{align}

\bibliography{grapheneHAdsorbates}
\end{document}